\documentclass[prb,twocolumn,superscriptaddress,amsmath,amssymb]{revtex4}
\usepackage{natbib}
\usepackage{graphicx}         
\usepackage{dcolumn}            
\usepackage{bm}               

\begin{document}

\title{Josephson supercurrent in Nb/InN-nanowire/Nb junctions}

\author{R. Frielinghaus}
\affiliation{Institute of Bio- and Nanosystems (IBN-1) and
JARA-Fundamentals of Future Information Technology,
Forschungszentrum J\"ulich, 52425 J\"ulich, Germany}

\author{I. E. Batov}
\affiliation{Institute of Solid State Physics, Russian Academy of
Sciences, Chernogolovka, Moscow district, Institutskaya 2, 142432
Russia}

\author{M. Weides}\altaffiliation{Current address: Physics Department University of
California Santa Barbara, 93106 Santa Barbara CA, USA}
\affiliation{Institute of Solid State Research (IFF) and
JARA-Fundamentals of Future Information Technology,
Forschungszentrum J\"ulich, 52425 J\"ulich, Germany}

\author{H. Kohlstedt} \altaffiliation{Nanoelektronik, Technische Fakult\"at,
Christian-Albrechts-Universit\"at zu Kiel, 24143 Kiel, Germany}
\affiliation{Institute of Solid State Research (IFF) and
JARA-Fundamentals of Future Information Technology,
Forschungszentrum J\"ulich, 52425 J\"ulich, Germany}

\author{R. Calarco}
\affiliation{Institute of Bio- and Nanosystems (IBN-1) and
JARA-Fundamentals of Future Information Technology,
Forschungszentrum J\"ulich, 52425 J\"ulich, Germany}

\author{Th. Sch\"apers}
\email{th.schaepers@fz-juelich.de} \affiliation{Institute of Bio-
and Nanosystems (IBN-1) and JARA-Fundamentals of Future
Information Technology, Forschungszentrum J\"ulich, 52425
J\"ulich, Germany}

\date{\today}

\hyphenation{InN}

\begin{abstract}
We experimentally studied the Josephson supercurrent in
Nb/InN-nanowire/Nb junctions. Large critical currents up to
5.7~$\mu$A have been achieved, which proves the good coupling of
the nanowire to the superconductor. The effect of a magnetic field
perpendicular to the plane of the Josephson junction on the
critical current has been studied. The observed monotonous
decrease of the critical current with magnetic field is explained
by the magnetic pair-breaking effect in planar Josephson junctions
of ultra-narrow width [J. C. Cuevas and F. S. Bergeret, Phys. Rev.
Lett.  99, 217002 (2007)]
\end{abstract}

\maketitle

Superconductor/normal-conductor/superconductor (SNS) junctions
with a semiconductor employed as the N-weak link material offer
the great advantage that here the Josephson supercurrent can be
controlled by means of the field
effect.\cite{Schaepers01a,Golubov04} Gate-controlled
superconductor/semiconductor hybrid devices such as
superconducting field effect transistors\cite{Akazaki96} or
split-gate structures\cite{Takayanagi95d} have been fabricated
which find no counterpart in conventional SNS structures. In
addition, the high carrier mobility attainable in semiconductors
in combination with the phase-coherent Andreev reflection leads to
novel unique phenomena in the
magnetotransport.\cite{Batov04,Eroms05,Batov07}  Usually for these
devices the semiconductor is patterned by conventional
lithography. As an elegant alternative one can also directly
create semiconductor nanostructures, i.e. nanowires, by epitaxial
growth.\cite{Thelander06} By using InAs nanowires connected to
superconducting electrodes tunable Josephson supercurrents,
supercurrent reversal, and Kondo-enhanced Andreev tunneling have
been realized.\cite{Doh05,vanDam06,Jespersen07}

Among the various materials used for semiconductor nanowires InN
is of particular interest for semiconductor/superconductor hybrid
structures, since the surface accumulation layer in InN can
provide a sufficiently low resistive contact to superconducting
electrodes.\cite{Chang05,Calarco07,Werner09} Due to almost ideal
crystalline properties of InN nanowires electronic transport along
the wires, contacted by normal metal electrodes, shows
quantization phenomena, i.e. flux periodic magnetoconductance
oscillations.\cite{Richter08} Furthermore, the carrier
concentration in the surface electron gas is of the order of
$10^{13}$~cm$^{-2}$ and thus about a factor of ten larger than in
InAs. Consequently when combined with superconducting electrodes
one can expect low resistive SNS junctions.

Here, we report on transport studies of Nb/InN-nanowire/Nb
junctions. We succeeded in observing a pronounced Josephson
supercurrent and a relatively large $I_cR_N$ product of up to
0.44~mV. The latter factor, the critical current times the normal
resistance, is an important figure of merit   for Josephson
junctions. We devoted special attention to the dependence of the
critical current $I_c$ on an external magnetic field $B$, where a
monotonous decrease of $I_c$ with $B$ was found. This experimental
finding is interpreted in the framework of a recent theoretical
model for the proximity effect in narrow-width junctions with
dimensions comparable or smaller than the magnetic length
$\xi_B=\sqrt{\Phi_0/B}$, where $\Phi_0=h/2e$ is the flux
quantum.\cite{Hammer07,Cuevas07}

\begin{figure}[]
\begin{center}
\includegraphics[width=1.0\columnwidth]{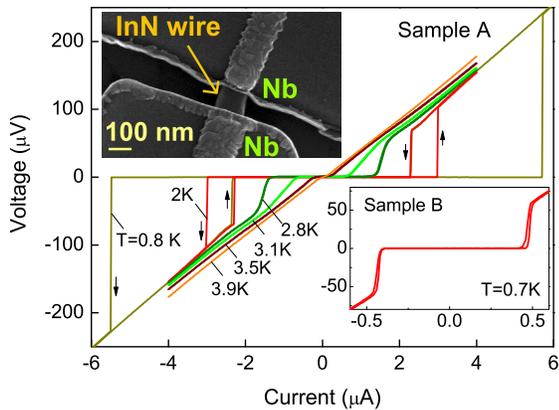}
\caption{(Color online) $I-V$ characteristics of sample~A at
various temperatures. The lower right inset shows the $I-V$
characteristics for sample~B at 0.7~K. The upper left inset shows
an scanning electron beam micrograph picture of sample A.}
\label{fig:1}
\end{center}
\end{figure}

The InN nanowires used for the normal conducting part of our
junctions were grown without catalyst on a Si (111) substrate by
plasma-assisted molecular beam epitaxy.\cite{Stoica06} The wires
had a typical length of 1~$\mu$m. The nanowires were contacted by
a pair of 100-nm-thick Nb electrodes. Before the Nb sputter
deposition the contact area was cleaned by Ar$^+$ milling. The
superconducting transition temperature $T_c$ of the Nb layers was
8.5~K. The InN nanowire of the first junction (sample A) had a
diameter $d=120$~nm and a Nb electrode separation $L=105$~nm [cf.
Fig.~\ref{fig:1} (inset)], while for the second junction (sample
B) the corresponding dimensions were 85~nm and 130~nm,
respectively. From measurements on back-gate transistor structures
performed on nanowires prepared in the same epitaxial run a
typical electron concentration of $1 \times 10^{19}$~cm$^{-3}$ was
determined. From measurements on nanowires contacted with normal
contacts with various distances a specific resistance of $\rho=4.2
\times 10^{-4}$~$\Omega$cm was estimated.\cite{Richter09} Using
these values we calculated a diffusion constant of
$\mathcal{D}=110$~cm$^{2}$/s.

The transport measurements were conducted in a \mbox{He-3}
cryostat in a temperature range from 0.7~K to 10~K. The magnetic
field was applied perpendicularly to the plane of the Nb
electrodes. The differential resistance was measured with a
lock-in amplifier by superimposing a small 17~Hz ac signal of
50~nA to the junction bias current.

The current-voltage ($I-V$) characteristics of sample~A for
various temperatures is shown in Fig.~\ref{fig:1}. As can be seen
here, a clear Josephson supercurrent is observed at temperatures
up to 3.5~K. At 0.8~K a critical current of 5.7~$\mu$A was
extracted. For temperatures below 2.5~K the $I-V$ characteristics
is hysteretic. The retrapping current $I_r$, characterized by the
switching from the voltage biased state back into the
superconducting state depends only slightly on temperature, with a
typical value of 2.2~$\mu$A. As can be seen in Fig.~\ref{fig:1}
(inset), for sample~B a lower critical current of 0.44~$\mu$A at
0.7~K was measured.

\begin{figure}[]
\begin{center}
\includegraphics[width=1.0\columnwidth]{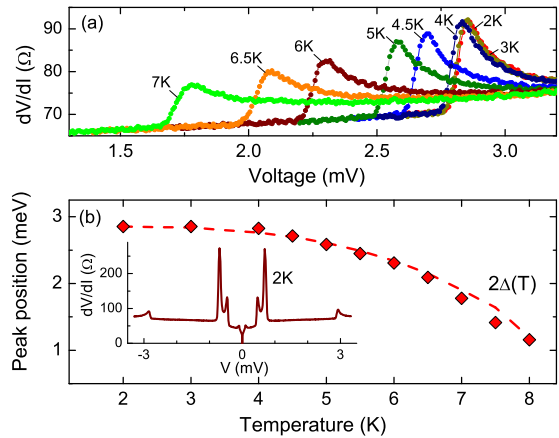}
\caption{(Color online) (a) Differential resistance $dV/dI$ of
sample~A as a function of bias voltage at various temperatures.
(b) Position of the peak assigned to $2\Delta$ as a function of
temperature. The broken line shows the expected value of 2$\Delta$
according to theory. The inset shows $dV/dI$ at 2~K in the full
bias voltage range.} \label{fig:2}
\end{center}
\end{figure}

The differential resistance $dV/dI$ as a function of the bias
voltage close to $2\Delta/e$ is shown in Fig.~\ref{fig:2}(a) for
temperatures in the range from 2 to 7~K. The distinct peak and the
lowering of $dV/dI$ at its lower bias side can be attributed to
the onset of multiple Andreev reflection.\cite{Octavio83,Aminov96}
The relatively small decrease of differential resistance at the
lower bias side of the peak by about 10\% compared to the higher
bias side can be attributed to the presence of an interface
barrier.\cite{Octavio83,Aminov96} As can be seen in
Fig.~\ref{fig:2}(b) (inset), more structures are found in the
differential resistance by approaching zero bias. Details about
these features, which we also attribute to multiple Andreev
reflection, will be given in a forthcoming publication. Further
evidence that the maxima shown in Fig.~\ref{fig:2}(a) can indeed
be assigned to the onset of Andreev reflection at $2\Delta/e$ is
given by the plot of the peak position as a function of
temperature [cf. Fig.~\ref{fig:2}(b)], since here the peak
position closely follows the theoretically expected decrease of
$2\Delta$ for an electron phonon-coupling strength of
$2\Delta_0/k_BT_c\simeq 3.9$.\cite{Carbotte90} Here, $\Delta_0$ is
the superconducting gap at $T=0$. At a temperature of 2~K and bias
voltages above $2\Delta/e$ a differential resistance of
78~$\Omega$ is measured. If this value is taken as the normal
state resistance $R_N$ of the junction one obtains a large
$I_cR_N$ product of 0.44~mV. For sample~B a normal state
resistance of 250~$\Omega$ was measured which results in a
somewhat lower $I_cR_N$ product of 0.11~mV.

\begin{figure}[]
\begin{center}
\includegraphics[width=1.0\columnwidth]{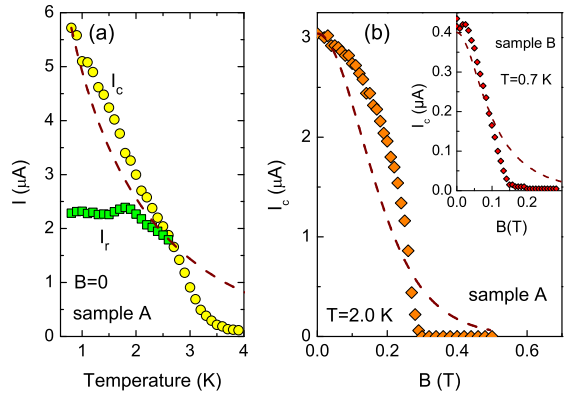}
\caption{(Color online) (a) $I_c$ ($\circ$) and $I_r$ ($\square$)
vs. $T$ of sample~A. The dashed line represents the calculated
values following Ref. [\onlinecite{Hammer07}]. (b) $I_c$ as a
function of $B$ of sample~A. The dashed line corresponds to the
calculated $I_c(B)$ dependence following
Ref.~[\onlinecite{Hammer07}]. The inset shows the corresponding
values for sample~B.} \label{fig:3}
\end{center}
\end{figure}

As can be seen in Fig.~\ref{fig:3}(a), the critical current $I_c$
of sample~A monotonously decreases with increasing temperature. A
complete suppression of the Josephson supercurrent is obtained at
about 3.7~K.  Up to 2~K the return current $I_r$ is almost
constant at a value of approximately 2.3~$\mu$A, while at higher
temperatures $T \geq 2.5$~K $I_r$ merges with $I_c$. A similar
behavior of the retrapping current was observed previously in
other Nb-semiconductor-Nb junctions.\cite{Schaepers97} As it was
recently pointed out by Courtois \emph{et al.},\cite{Courtois08}
the hysteresis in the $I-V$ characteristics of proximity SNS
structures can be attributed to the increase of the normal-metal
electron temperature once the junction switches to the resistive
state.

From the transport data of the InN nanowires one extracts an
elastic mean free path of approximately 45~nm, thus the transport
takes place in the diffusive regime. In addition, as stated above
we have to consider the presence of an interface barrier. For this
case, the critical current was studied theoretically by Hammer
\emph{et al.}\cite{Hammer07} In Fig.~\ref{fig:3}(a) the
corresponding theoretical curve which fits best to the
experimental values is plotted. We followed the approach of Dubos
\emph{et al.}\cite{Dubos01} and Carillo \emph{et
al.}\cite{Carillo08} by using a reduced effective Thouless energy
$E^*_{Th}$ as a fitting parameter. The lower value of
$E^*_{Th}=0.15$~meV compared to $E_{Th}=\hbar
\mathcal{D}/L^2=0.67$~meV obtained from the transport parameters
is a measure of the detrimental effect of the interface
resistance.\cite{Hammer07}

As can be seen in Fig.~\ref{fig:3}(b), $I_c(B)$ of sample~A
monotonously decreases with increasing magnetic field. For
magnetic fields larger than 0.3~T the Josephson supercurrent is
completely suppressed. In contrast to wide S/semiconductor/S
Josephson junctions,\cite{Neurohr99} no Fraunhofer-type
interference pattern of $I_c(B)$ is observed. The absence of a
magnetic interference pattern in SNS structures was first observed
by Anger \emph{et al.}\cite{Angers08} and theoretically explained
by Hammer \emph{et al.}\cite{Hammer07} and Cuevas and
Bergeret.\cite{Cuevas07} The reason for the monotonous decay of
$I_c$ is that for junctions with a width smaller than the magnetic
length $\xi_B$ the magnetic field acts as a pair-breaking factor.
Indeed at the field of 0.16~T where the first minimum at $\Phi_0$
is expected in the Fraunhofer interference pattern the magnetic
length $\xi_B$ is as large as 110~nm and thus comparable to the
junction width. For sample~B a similar dependence of $I_c$ on $B$
is observed with a full suppression of $I_c$ at 0.2~T. By using
the model of Hammer \emph{et al.}\cite{Hammer07} for the the case
of low transparent junctions we calculated the expected dependence
of $I_c$ on $B$ for  $E^*_{Th}= 0.15$~meV. As can be seen in
Fig.~\ref{fig:3}(b), a reasonable agreement between experiment and
theory is obtained. The same is true for sample~B with
$E^*_{Th}=0.7$~meV [cf. Fig.~\ref{fig:3}(b), inset]. A possible
reason for the discrepancy between the experimental values and
theoretical curves might be that in our InN nanowires the current
flows mainly in the surface accumulation layer, which leads to an
inhomogeneous current distribution.

In summary, superconducting Nb/InN-nanowire/Nb junctions with
large critical currents up to 5.7~$\mu$A and large $I_cR_N$
products up to 0.44~mV have been fabricated. Owing to the small
width of nanowires a monotonous decrease of $I_c$ with $B$ was
observed, since in this case the magnetic field is the main pair
breaking factor. The present results suggest that
Nb/InN-nanowire/Nb structures are well suited for fundamental
research and application in nano-scaled Josephson junction-based
devices.

We are grateful to A. A. Golubov (Twente University, The
Netherlands) and V. V. Ryazanov (Institute of Solid State Physics
RAS, Chernogolovka) for fruitful discussions and H. Kertz for
support during the measurements. I.E.B. acknowledges the Russian
Foundation for Basic Research: project RFBR 09-02-01499 for
financial support.


\begin{thebibliography}{27}
\expandafter\ifx\csname
natexlab\endcsname\relax\def\natexlab#1{#1}\fi
\expandafter\ifx\csname bibnamefont\endcsname\relax
  \def\bibnamefont#1{#1}\fi
\expandafter\ifx\csname bibfnamefont\endcsname\relax
  \def\bibfnamefont#1{#1}\fi
\expandafter\ifx\csname citenamefont\endcsname\relax
  \def\citenamefont#1{#1}\fi
\expandafter\ifx\csname url\endcsname\relax
  \def\url#1{\texttt{#1}}\fi
\expandafter\ifx\csname
urlprefix\endcsname\relax\def\urlprefix{URL }\fi
\providecommand{\bibinfo}[2]{#2}
\providecommand{\eprint}[2][]{\url{#2}}

\bibitem[{\citenamefont{Golubov et~al.}(2004)\citenamefont{Golubov, Kupriyanov,
  and Il'ichev}}]{Golubov04}
\bibinfo{author}{\bibfnamefont{A.~A.} \bibnamefont{Golubov}},
  \bibinfo{author}{\bibfnamefont{M.~Y.} \bibnamefont{Kupriyanov}},
  \bibnamefont{and} \bibinfo{author}{\bibfnamefont{E.}~\bibnamefont{Il'ichev}},
  \bibinfo{journal}{Rev. Mod. Phys.} \textbf{\bibinfo{volume}{76}},
  \bibinfo{eid}{411} (\bibinfo{year}{2004}).

\bibitem[{\citenamefont{\mbox{T}. Sch\"apers}(2001)}]{Schaepers01a}
\bibinfo{author}{\bibnamefont{\mbox{T}. Sch\"apers}},
  \emph{\bibinfo{title}{{S}uperconductor/{S}emiconductor {J}unctions}}, vol.
  \bibinfo{volume}{174} of \emph{\bibinfo{series}{Springer Tracts on Modern
  Physics}} (\bibinfo{publisher}{Springer-Verlag, Berlin Heidelberg},
  \bibinfo{year}{2001}).

\bibitem[{\citenamefont{Akazaki et~al.}(1996)\citenamefont{Akazaki, Takayanagi,
  Nitta, and Enoki}}]{Akazaki96}
\bibinfo{author}{\bibfnamefont{T.}~\bibnamefont{Akazaki}},
  \bibinfo{author}{\bibfnamefont{H.}~\bibnamefont{Takayanagi}},
  \bibinfo{author}{\bibfnamefont{J.}~\bibnamefont{Nitta}}, \bibnamefont{and}
  \bibinfo{author}{\bibfnamefont{T.}~\bibnamefont{Enoki}},
  \bibinfo{journal}{Appl. Phys. Lett.} \textbf{\bibinfo{volume}{68}},
  \bibinfo{pages}{418} (\bibinfo{year}{1996}).

\bibitem[{\citenamefont{Takayanagi et~al.}(1995)\citenamefont{Takayanagi,
  Akazaki, and Nitta}}]{Takayanagi95d}
\bibinfo{author}{\bibfnamefont{H.}~\bibnamefont{Takayanagi}},
  \bibinfo{author}{\bibfnamefont{T.}~\bibnamefont{Akazaki}}, \bibnamefont{and}
  \bibinfo{author}{\bibfnamefont{J.}~\bibnamefont{Nitta}},
  \bibinfo{journal}{Phys. Rev. Lett.} \textbf{\bibinfo{volume}{75}},
  \bibinfo{pages}{3533} (\bibinfo{year}{1995}).

\bibitem[{\citenamefont{Batov et~al.}(2004)\citenamefont{Batov, \mbox{Th}.
  Sch\"apers, Golubov, and Ustinov}}]{Batov04}
\bibinfo{author}{\bibfnamefont{I.~E.} \bibnamefont{Batov}},
  \bibinfo{author}{\bibnamefont{\mbox{Th}. Sch\"apers}},
  \bibinfo{author}{\bibfnamefont{A.~A.} \bibnamefont{Golubov}},
  \bibnamefont{and} \bibinfo{author}{\bibfnamefont{A.~V.}
  \bibnamefont{Ustinov}}, \bibinfo{journal}{J. Appl. Phys.}
  \textbf{\bibinfo{volume}{96}}, \bibinfo{pages}{3366} (\bibinfo{year}{2004}).

\bibitem[{\citenamefont{Eroms et~al.}(2005)\citenamefont{Eroms, Weiss, Boeck,
  Borghs, and Z\"ulicke}}]{Eroms05}
\bibinfo{author}{\bibfnamefont{J.}~\bibnamefont{Eroms}},
  \bibinfo{author}{\bibfnamefont{D.}~\bibnamefont{Weiss}},
  \bibinfo{author}{\bibfnamefont{J.~D.} \bibnamefont{Boeck}},
  \bibinfo{author}{\bibfnamefont{G.}~\bibnamefont{Borghs}}, \bibnamefont{and}
  \bibinfo{author}{\bibfnamefont{U.}~\bibnamefont{Z\"ulicke}},
  \bibinfo{journal}{Phys. Rev. Lett.} \textbf{\bibinfo{volume}{95}},
  \bibinfo{pages}{107001/1} (\bibinfo{year}{2005}).

\bibitem[{\citenamefont{Batov et~al.}(2007)\citenamefont{Batov, Sch\"{a}pers,
  Chtchelkatchev, Hardtdegen, and Ustinov}}]{Batov07}
\bibinfo{author}{\bibfnamefont{I.~E.} \bibnamefont{Batov}},
  \bibinfo{author}{\bibfnamefont{T.}~\bibnamefont{Sch\"{a}pers}},
  \bibinfo{author}{\bibfnamefont{N.~M.} \bibnamefont{Chtchelkatchev}},
  \bibinfo{author}{\bibfnamefont{H.}~\bibnamefont{Hardtdegen}},
  \bibnamefont{and} \bibinfo{author}{\bibfnamefont{A.~V.}
  \bibnamefont{Ustinov}}, \bibinfo{journal}{Phys. Rev. B}
  \textbf{\bibinfo{volume}{76}}, \bibinfo{eid}{115313}
  (\bibinfo{year}{2007}).

\bibitem[{\citenamefont{Thelander et~al.}(2006)\citenamefont{Thelander,
  Agarwal, Brongersma, Eymery, Feiner, Forchel, Scheffler, Riess, Ohlsson,
  G\"osele et~al.}}]{Thelander06}
\bibinfo{author}{\bibfnamefont{C.}~\bibnamefont{Thelander}},
  \bibinfo{author}{\bibfnamefont{P.}~\bibnamefont{Agarwal}},
  \bibinfo{author}{\bibfnamefont{S.}~\bibnamefont{Brongersma}},
  \bibinfo{author}{\bibfnamefont{J.}~\bibnamefont{Eymery}},
  \bibinfo{author}{\bibfnamefont{L.}~\bibnamefont{Feiner}},
  \bibinfo{author}{\bibfnamefont{A.}~\bibnamefont{Forchel}},
  \bibinfo{author}{\bibfnamefont{M.}~\bibnamefont{Scheffler}},
  \bibinfo{author}{\bibfnamefont{W.}~\bibnamefont{Riess}},
  \bibinfo{author}{\bibfnamefont{B.}~\bibnamefont{Ohlsson}},
  \bibinfo{author}{\bibfnamefont{U.}~\bibnamefont{G\"osele}},
  \bibnamefont{et~al.}, \bibinfo{journal}{Materials Today}
  \textbf{\bibinfo{volume}{9}}, \bibinfo{pages}{28} (\bibinfo{year}{2006}).

\bibitem[{\citenamefont{van Dam et~al.}(2006)\citenamefont{van Dam, Nazarov,
  Bakkers, Franceschi, and Kouwenhoven}}]{vanDam06}
\bibinfo{author}{\bibfnamefont{J.~A.} \bibnamefont{van Dam}},
  \bibinfo{author}{\bibfnamefont{Y.~V.} \bibnamefont{Nazarov}},
  \bibinfo{author}{\bibfnamefont{E.~P. A.~M.} \bibnamefont{Bakkers}},
  \bibinfo{author}{\bibfnamefont{S.~D.} \bibnamefont{Franceschi}},
  \bibnamefont{and} \bibinfo{author}{\bibfnamefont{L.~P.}
  \bibnamefont{Kouwenhoven}}, \bibinfo{journal}{Nature}
  \textbf{\bibinfo{volume}{442}}, \bibinfo{pages}{667} (\bibinfo{year}{2006}).

\bibitem[{\citenamefont{Doh et~al.}(2005)\citenamefont{Doh, van Dam, Roest,
  Bakkers, Kouwenhoven, and Franceschi}}]{Doh05}
\bibinfo{author}{\bibfnamefont{Y.-J.} \bibnamefont{Doh}},
  \bibinfo{author}{\bibfnamefont{J.~A.} \bibnamefont{van Dam}},
  \bibinfo{author}{\bibfnamefont{A.~L.} \bibnamefont{Roest}},
  \bibinfo{author}{\bibfnamefont{E.~P. A.~M.} \bibnamefont{Bakkers}},
  \bibinfo{author}{\bibfnamefont{L.~P.} \bibnamefont{Kouwenhoven}},
  \bibnamefont{and} \bibinfo{author}{\bibfnamefont{S.~D.}
  \bibnamefont{Franceschi}}, \bibinfo{journal}{Science}
  \textbf{\bibinfo{volume}{309}}, \bibinfo{pages}{272} (\bibinfo{year}{2005}).

\bibitem[{\citenamefont{Sand-Jespersen
  et~al.}(2007)\citenamefont{Sand-Jespersen, Paaske, Andersen, Grove-Rasmussen,
  rgensen, Aagesen, rensen, Lindelof, Flensberg, and rd}}]{Jespersen07}
\bibinfo{author}{\bibfnamefont{T.}~\bibnamefont{Sand-Jespersen}},
  \bibinfo{author}{\bibfnamefont{J.}~\bibnamefont{Paaske}},
  \bibinfo{author}{\bibfnamefont{B.~M.} \bibnamefont{Andersen}},
  \bibinfo{author}{\bibfnamefont{K.}~\bibnamefont{Grove-Rasmussen}},
  \bibinfo{author}{\bibfnamefont{H.~I.} \bibnamefont{J{\o}rgensen}},
  \bibinfo{author}{\bibfnamefont{M.}~\bibnamefont{Aagesen}},
  \bibinfo{author}{\bibfnamefont{C.~B.} \bibnamefont{S{\o}rensen}},
  \bibinfo{author}{\bibfnamefont{P.~E.} \bibnamefont{Lindelof}},
  \bibinfo{author}{\bibfnamefont{K.}~\bibnamefont{Flensberg}},
  \bibnamefont{and} \bibinfo{author}{\bibfnamefont{J.} \bibnamefont{Nyg{\aa}rd}},
  \bibinfo{journal}{Phys. Rev. Lett.} \textbf{\bibinfo{volume}{99}},
  \bibinfo{eid}{126603} (\bibinfo{year}{2007}),.

\bibitem[{\citenamefont{Chang et~al.}(2005)\citenamefont{Chang, Chi, Wang,
  Chen, Chen, Ren, and Pearton}}]{Chang05}
\bibinfo{author}{\bibfnamefont{C.-Y.} \bibnamefont{Chang}},
  \bibinfo{author}{\bibfnamefont{G.-C.} \bibnamefont{Chi}},
  \bibinfo{author}{\bibfnamefont{W.-M.} \bibnamefont{Wang}},
  \bibinfo{author}{\bibfnamefont{L.-C.} \bibnamefont{Chen}},
  \bibinfo{author}{\bibfnamefont{K.-H.} \bibnamefont{Chen}},
  \bibinfo{author}{\bibfnamefont{F.}~\bibnamefont{Ren}}, \bibnamefont{and}
  \bibinfo{author}{\bibfnamefont{S.~J.} \bibnamefont{Pearton}},
  \bibinfo{journal}{Appl. Phys. Lett.} \textbf{\bibinfo{volume}{87}},
  \bibinfo{pages}{093112} (\bibinfo{year}{2005}).

\bibitem[{\citenamefont{Werner et~al.}(2009)\citenamefont{Werner, Limbach,
  Carsten, Denker, Malindretos, and Rizzi}}]{Werner09}
\bibinfo{author}{\bibfnamefont{F.}~\bibnamefont{Werner}},
  \bibinfo{author}{\bibfnamefont{F.}~\bibnamefont{Limbach}},
  \bibinfo{author}{\bibfnamefont{M.}~\bibnamefont{Carsten}},
  \bibinfo{author}{\bibfnamefont{C.}~\bibnamefont{Denker}},
  \bibinfo{author}{\bibfnamefont{J.}~\bibnamefont{Malindretos}},
  \bibnamefont{and} \bibinfo{author}{\bibfnamefont{A.}~\bibnamefont{Rizzi}},
  \bibinfo{journal}{Nano Lett.} \textbf{\bibinfo{volume}{9}},
  \bibinfo{pages}{1567} (\bibinfo{year}{2009}).

\bibitem[{\citenamefont{Calarco and Marso}(2007)}]{Calarco07}
\bibinfo{author}{\bibfnamefont{R.}~\bibnamefont{Calarco}} \bibnamefont{and}
  \bibinfo{author}{\bibfnamefont{M.}~\bibnamefont{Marso}},
  \bibinfo{journal}{Appl. Phys. A} \textbf{\bibinfo{volume}{87}},
  \bibinfo{pages}{499} (\bibinfo{year}{2007}).

\bibitem[{\citenamefont{Richter et~al.}(2008)\citenamefont{Richter, \mbox{Ch.}
  Bl\"omers, L\"uth, Calarco, Indlekofer, Marso, and \mbox{Th.}
  Sch\"apers}}]{Richter08}
\bibinfo{author}{\bibfnamefont{T.}~\bibnamefont{Richter}},
  \bibinfo{author}{\bibnamefont{\mbox{Ch.} Bl\"omers}},
  \bibinfo{author}{\bibfnamefont{H.}~\bibnamefont{L\"uth}},
  \bibinfo{author}{\bibfnamefont{R.}~\bibnamefont{Calarco}},
  \bibinfo{author}{\bibfnamefont{M.}~\bibnamefont{Indlekofer}},
  \bibinfo{author}{\bibfnamefont{M.}~\bibnamefont{Marso}}, \bibnamefont{and}
  \bibinfo{author}{\bibnamefont{\mbox{Th.} Sch\"apers}}, \bibinfo{journal}{Nano
  Lett.} \textbf{\bibinfo{volume}{8}}, \bibinfo{pages}{2834}
  (\bibinfo{year}{2008}).

\bibitem[{\citenamefont{Hammer et~al.}(2007)\citenamefont{Hammer, Cuevas,
  Bergeret, and Belzig}}]{Hammer07}
\bibinfo{author}{\bibfnamefont{J.~C.} \bibnamefont{Hammer}},
  \bibinfo{author}{\bibfnamefont{J.~C.} \bibnamefont{Cuevas}},
  \bibinfo{author}{\bibfnamefont{F.~S.} \bibnamefont{Bergeret}},
  \bibnamefont{and} \bibinfo{author}{\bibfnamefont{W.}~\bibnamefont{Belzig}},
  \bibinfo{journal}{Phys. Rev. B}
  \textbf{\bibinfo{volume}{76}}, \bibinfo{eid}{064514}
  (\bibinfo{year}{2007}).

\bibitem[{\citenamefont{Cuevas and Bergeret}(2007)}]{Cuevas07}
\bibinfo{author}{\bibfnamefont{J.~C.} \bibnamefont{Cuevas}} \bibnamefont{and}
  \bibinfo{author}{\bibfnamefont{F.~S.} \bibnamefont{Bergeret}},
  \bibinfo{journal}{Phys. Rev. Lett.} \textbf{\bibinfo{volume}{99}},
  \bibinfo{eid}{217002}  (\bibinfo{year}{2007}).

\bibitem[{\citenamefont{Stoica et~al.}(2006)\citenamefont{Stoica, Meijers,
  Calarco, Richter, and L\"uth}}]{Stoica06}
\bibinfo{author}{\bibfnamefont{T.}~\bibnamefont{Stoica}},
  \bibinfo{author}{\bibfnamefont{R.}~\bibnamefont{Meijers}},
  \bibinfo{author}{\bibfnamefont{R.}~\bibnamefont{Calarco}},
  \bibinfo{author}{\bibfnamefont{T.}~\bibnamefont{Richter}}, \bibnamefont{and}
  \bibinfo{author}{\bibfnamefont{H.}~\bibnamefont{L\"uth}},
  \bibinfo{journal}{J. Cryst. Growth} \textbf{\bibinfo{volume}{290}},
  \bibinfo{pages}{241} (\bibinfo{year}{2006}).

\bibitem[{\citenamefont{Richter et~al.}(2009)\citenamefont{Richter, L\"uth,
  Sch\"apers, Meijers, Jeganathan, Hernandez, Calarco, and Marso}}]{Richter09}
\bibinfo{author}{\bibfnamefont{T.}~\bibnamefont{Richter}},
  \bibinfo{author}{\bibfnamefont{H.}~\bibnamefont{L\"uth}},
  \bibinfo{author}{\bibfnamefont{T.}~\bibnamefont{Sch\"apers}},
  \bibinfo{author}{\bibfnamefont{R.}~\bibnamefont{Meijers}},
  \bibinfo{author}{\bibfnamefont{K.}~\bibnamefont{Jeganathan}},
  \bibinfo{author}{\bibfnamefont{S.~E.} \bibnamefont{Hernandez}},
  \bibinfo{author}{\bibfnamefont{R.}~\bibnamefont{Calarco}}, \bibnamefont{and}
  \bibinfo{author}{\bibfnamefont{M.}~\bibnamefont{Marso}},
  \bibinfo{journal}{Nanotechnol.} \textbf{\bibinfo{volume}{20}},
  \bibinfo{pages}{405206} (\bibinfo{year}{2009}).

\bibitem[{\citenamefont{Octavio et~al.}(1983)\citenamefont{Octavio, Tinkham,
  Blonder, and Klapwijk}}]{Octavio83}
\bibinfo{author}{\bibfnamefont{M.}~\bibnamefont{Octavio}},
  \bibinfo{author}{\bibfnamefont{M.}~\bibnamefont{Tinkham}},
  \bibinfo{author}{\bibfnamefont{G.~E.} \bibnamefont{Blonder}},
  \bibnamefont{and} \bibinfo{author}{\bibfnamefont{T.~M.}
  \bibnamefont{Klapwijk}}, \bibinfo{journal}{Phys. Rev. B}
  \textbf{\bibinfo{volume}{27}}, \bibinfo{pages}{6739} (\bibinfo{year}{1983}).

\bibitem[{\citenamefont{Aminov et~al.}(1996)\citenamefont{Aminov, Golubov, and
  \mbox{Yu}. Kupriyanov}}]{Aminov96}
\bibinfo{author}{\bibfnamefont{B.~A.} \bibnamefont{Aminov}},
  \bibinfo{author}{\bibfnamefont{A.~A.} \bibnamefont{Golubov}},
  \bibnamefont{and} \bibinfo{author}{\bibfnamefont{M.}~\bibnamefont{\mbox{Yu}.
  Kupriyanov}}, \bibinfo{journal}{Phys. Rev. B} \textbf{\bibinfo{volume}{53}},
  \bibinfo{pages}{365} (\bibinfo{year}{1996}).

\bibitem{Carbotte90}
\bibinfo{author}{\bibfnamefont{J.~P.} \bibnamefont{Carbotte}},
  \bibinfo{journal}{Rev. Mod. Phys.} \textbf{\bibinfo{volume}{62}},
  \bibinfo{pages}{1027} (\bibinfo{year}{1990}) and references therein.

\bibitem[{\citenamefont{\mbox{Th}. Sch\"apers
  et~al.}(1997)\citenamefont{\mbox{Th}. Sch\"apers, Kaluza, Neurohr,
  Malindretos, Crecelius, van~der Hart, Hardtdegen, and L\"uth}}]{Schaepers97}
\bibinfo{author}{\bibnamefont{\mbox{Th}. Sch\"apers}},
  \bibinfo{author}{\bibfnamefont{A.}~\bibnamefont{Kaluza}},
  \bibinfo{author}{\bibfnamefont{K.}~\bibnamefont{Neurohr}},
  \bibinfo{author}{\bibfnamefont{J.}~\bibnamefont{Malindretos}},
  \bibinfo{author}{\bibfnamefont{G.}~\bibnamefont{Crecelius}},
  \bibinfo{author}{\bibfnamefont{A.}~\bibnamefont{van~der Hart}},
  \bibinfo{author}{\bibfnamefont{H.}~\bibnamefont{Hardtdegen}},
  \bibnamefont{and} \bibinfo{author}{\bibfnamefont{H.}~\bibnamefont{L\"uth}},
  \bibinfo{journal}{Appl. Phys. Lett.} \textbf{\bibinfo{volume}{71}},
  \bibinfo{pages}{3537} (\bibinfo{year}{1997}).

\bibitem[{\citenamefont{Courtois et~al.}(2008)\citenamefont{Courtois, Meschke,
  Peltonen, and Pekola}}]{Courtois08}
\bibinfo{author}{\bibfnamefont{H.}~\bibnamefont{Courtois}},
  \bibinfo{author}{\bibfnamefont{M.}~\bibnamefont{Meschke}},
  \bibinfo{author}{\bibfnamefont{J.~T.} \bibnamefont{Peltonen}},
  \bibnamefont{and} \bibinfo{author}{\bibfnamefont{J.~P.}
  \bibnamefont{Pekola}}, \bibinfo{journal}{Physical Review Letters}
  \textbf{\bibinfo{volume}{101}}, \bibinfo{eid}{067002}
  (\bibinfo{year}{2008}).

\bibitem[{\citenamefont{Dubos et~al.}(2001)\citenamefont{Dubos, Courtois,
  Pannetier, Wilhelm, Zaikin, and Sch\"on}}]{Dubos01}
\bibinfo{author}{\bibfnamefont{P.}~\bibnamefont{Dubos}},
  \bibinfo{author}{\bibfnamefont{H.}~\bibnamefont{Courtois}},
  \bibinfo{author}{\bibfnamefont{B.}~\bibnamefont{Pannetier}},
  \bibinfo{author}{\bibfnamefont{F.~K.} \bibnamefont{Wilhelm}},
  \bibinfo{author}{\bibfnamefont{A.~D.} \bibnamefont{Zaikin}},
  \bibnamefont{and} \bibinfo{author}{\bibfnamefont{G.}~\bibnamefont{Sch\"on}},
  \bibinfo{journal}{Phys. Rev. B} \textbf{\bibinfo{volume}{63}},
  \bibinfo{pages}{064502} (\bibinfo{year}{2001}).

\bibitem[{\citenamefont{Carillo et~al.}(2008)\citenamefont{Carillo, Born,
  Pellegrini, Tafuri, Biasiol, Sorba, and Beltram}}]{Carillo08}
\bibinfo{author}{\bibfnamefont{F.}~\bibnamefont{Carillo}},
  \bibinfo{author}{\bibfnamefont{D.}~\bibnamefont{Born}},
  \bibinfo{author}{\bibfnamefont{V.}~\bibnamefont{Pellegrini}},
  \bibinfo{author}{\bibfnamefont{F.}~\bibnamefont{Tafuri}},
  \bibinfo{author}{\bibfnamefont{G.}~\bibnamefont{Biasiol}},
  \bibinfo{author}{\bibfnamefont{L.}~\bibnamefont{Sorba}}, \bibnamefont{and}
  \bibinfo{author}{\bibfnamefont{F.}~\bibnamefont{Beltram}},
  \bibinfo{journal}{Phys. Rev. B}
  \textbf{\bibinfo{volume}{78}}, \bibinfo{eid}{052506}
  (\bibinfo{year}{2008}).

\bibitem[{\citenamefont{Neurohr et~al.}(1999)\citenamefont{Neurohr, \mbox{Th}.
  Sch\"apers, Malindretos, Lachenmann, Braginski, L\"uth, Behet, Borghs, and
  Golubov}}]{Neurohr99}
\bibinfo{author}{\bibfnamefont{K.}~\bibnamefont{Neurohr}},
  \bibinfo{author}{\bibnamefont{\mbox{Th}. Sch\"apers}},
  \bibinfo{author}{\bibfnamefont{J.}~\bibnamefont{Malindretos}},
  \bibinfo{author}{\bibfnamefont{S.}~\bibnamefont{Lachenmann}},
  \bibinfo{author}{\bibfnamefont{A.~I.} \bibnamefont{Braginski}},
  \bibinfo{author}{\bibfnamefont{H.}~\bibnamefont{L\"uth}},
  \bibinfo{author}{\bibfnamefont{M.}~\bibnamefont{Behet}},
  \bibinfo{author}{\bibfnamefont{G.}~\bibnamefont{Borghs}}, \bibnamefont{and}
  \bibinfo{author}{\bibfnamefont{A.~A.} \bibnamefont{Golubov}},
  \bibinfo{journal}{Phys. Rev. B} \textbf{\bibinfo{volume}{59}},
  \bibinfo{pages}{11197} (\bibinfo{year}{1999}).

\bibitem[{\citenamefont{Angers et~al.}(2008)\citenamefont{Angers, Chiodi,
  Montambaux, Ferrier, Gu\'{e}ron, Bouchiat, and Cuevas}}]{Angers08}
\bibinfo{author}{\bibfnamefont{L.}~\bibnamefont{Angers}},
  \bibinfo{author}{\bibfnamefont{F.}~\bibnamefont{Chiodi}},
  \bibinfo{author}{\bibfnamefont{G.}~\bibnamefont{Montambaux}},
  \bibinfo{author}{\bibfnamefont{M.}~\bibnamefont{Ferrier}},
  \bibinfo{author}{\bibfnamefont{S.}~\bibnamefont{Gu\'{e}ron}},
  \bibinfo{author}{\bibfnamefont{H.}~\bibnamefont{Bouchiat}}, \bibnamefont{and}
  \bibinfo{author}{\bibfnamefont{J.~C.} \bibnamefont{Cuevas}},
  \bibinfo{journal}{Phys. Rev. B}
  \textbf{\bibinfo{volume}{77}}, \bibinfo{eid}{165408}
 (\bibinfo{year}{2008}).

\end{thebibliography}

\newpage

\end{document}